\documentclass[aps, prb, twocolumn, superscriptaddress, showpacs]{revtex4}
\usepackage{graphicx,amsmath}
\usepackage[dvipdfm,colorlinks=true, citecolor=blue, urlcolor=blue ]{hyperref}
\renewcommand{\section}[1]{{\par\it #1.---}\ignorespaces}

\begin{document}

\title{Generating Many Majorana Modes via Periodic Driving: A Superconductor
Model}
\author{Qing-Jun Tong}
\affiliation{Center for Interdisciplinary Studies $\&$ Key Laboratory for Magnetism and
Magnetic Materials of the MoE, Lanzhou University, Lanzhou 730000, China}
\author{Jun-Hong An}
\email{anjhong@lzu.edu.cn}
\affiliation{Center for Interdisciplinary Studies $\&$ Key Laboratory for Magnetism and
Magnetic Materials of the MoE, Lanzhou University, Lanzhou 730000, China}
\affiliation{Center for Quantum Technologies and Department of Physics, National
University of Singapore, 3 Science Drive 2, Singapore 117543, Singapore}
\author{Jiangbin Gong}
\affiliation{Department of Physics and Centre for Computational Science and Engineering,
National University of Singapore, Singapore 117542, Singapore}
\author{Hong-Gang Luo}
\affiliation{Center for Interdisciplinary Studies $\&$ Key Laboratory for Magnetism and
Magnetic Materials of the MoE, Lanzhou University, Lanzhou 730000, China}
\affiliation{Beijing Computational Science Research Center, Beijing 100084, China}
\author{C. H. Oh}\email{phyohch@nus.edu.sg}
\affiliation{Center for Quantum Technologies and Department of Physics, National
University of Singapore, 3 Science Drive 2, Singapore 117543, Singapore}

\begin{abstract}
Realizing Majorana modes (MMs) in condensed-matter systems is of
vast experimental and theoretical {interests}, and some signatures
of MMs have been measured already. To facilitate future experimental
observations and to explore further applications of MMs, generating
\textit{many} MMs at ease in an experimentally accessible manner has
become one important issue. This task is achieved here in a
one-dimensional $p$-wave superconductor system with the nearest- and
next-nearest-neighbor interactions. In particular, a periodic
modulation of some system parameters can induce an effective
long-range interaction (as suggested by the Baker-Campbell-Hausdorff
formula) and may recover time-reversal symmetry already broken in
undriven cases. By exploiting these two independent mechanisms at
once we have established a general method in generating many Floquet
MMs via periodic driving.
\end{abstract}

\maketitle

\section{Introduction}
The Majorana fermion, a particle which is its own anti-particle, \cite%
{Majorana1937} is attracting tremendous attention. \cite%
{Nayak2008,Wilczek2009,Beenakker2011,Alicea2012} In addition to its
fundamental interest,
\cite{Moore1991,Read2000,Ivanov2001,Alicea2011} its potential
applications in topological quantum computation are also noteworthy.
\cite{Kitaev2003} Along with considerable theoretical studies,
\cite{Read2000,Kitaev2003,Fu2008,Sau2010,Alicea2010,Lutchyn2010,Oreg2010,Zhang2008,Sato2009,Zhu2011,Gong2012,Fidkowski2011}
the experimental search for Majorana modes (MMs) in condensed-matter
systems has become a timely and important research topic. Indeed,
following the theoretical results in Refs.~\onlinecite{Lutchyn2010,
Oreg2010, Sengupta2001,Law2009,Flensberg2010,Sau20101,Wimmer2011},
the zero-bias
conductance peaks observed recently \cite%
{Mourik2012,Williams2012,Deng2012,Ronen2012} are regarded as a
signature of MMs in one-dimensional (1D) spin-orbit coupled
semiconductor nanowires. However, the observed zero-bias peaks can
be due to other reasons as well, e.g., the strong disorder in the
nanowire \cite{Liu2012,Pikulin2012} or smooth confinement potential
at the wire end. \cite{Kells2012} This being the case, the formation
of MMs in these systems are yet to be double-confirmed by other
approaches.

To identify MMs and facilitate their experimental observation, the
signal strength should be
enhanced.~\cite{Fulga2011,Kraus2012,Wong2011} If \textit{many} MMs
are present at the same edge and topologically protected from
hybridizing with each other, one may verify if the signal originates
from MMs by tuning the actual number of them, with the enhanced
signal also more robust against experimental disorder
\cite{Flensberg2010,Liu2012,Shivamoggi2010} and contaminations from
thermal excitations.
\cite{Sengupta2001,Law2009,Flensberg2010,Sau20101,Wimmer2011} It is
thus constructive to find a general method to form many MMs within
one single system. In this respect, two facts are known. First, the
formation of many MMs needs the
protection of time-reversal symmetry.  \cite%
{Schnyder2008,Niu2012,Wong2011} Second, a longer-range interaction
in a system is helpful to obtain more than two pairs of MMs.
\cite{Niu2012} As such, the generation of many MMs is equivalent to
the following theoretical question: how to synthesize a long-range
interaction in a topologically nontrivial condensed-matter system
while maintaining time-reversal symmetry?

As a conceptual advance, our answer to this question is rather
simple and general. Given that periodic driving has become one
highly controllable and versatile tool in generating different
topological states of matter,
\cite{Oka2009,Inoue2010,Kitagawa20101,Kitagawa2010,Lindner2011,Jiang2011,Meidan2011,Fulga2012,Reynoso}
we show that a periodic driving protocol can create many Floquet MMs
because it can generically induce effective long-range interactions
and may also restore time-reversal symmetry (if it is broken without
driving). Note that Floquet MMs are a particular class of MMs
associated with the Floquet quasi-energy bands of a periodically
driven system:~\cite{Jiang2011} they may be used for topological
quantum computation as ``normal" MMs do. \cite{Floquet2012}

Specifically, we propose to generate multiple Floquet MMs by switching (periodic quenching) a
Hamiltonian from $H_1$
for the first half-period to $H_2$ for the second one. The Floquet operator $%
U$ is then
\begin{eqnarray}
U(T)=e^{-\frac{iH_2 T}{2\hbar}} e^{-\frac{i H_1 T}{2\hbar}} \equiv e^{-\frac{%
iH_\text{eff} T}{\hbar}},  \label{bf}
\end{eqnarray}
where an effective Hamiltonian $H_\text{eff}$ for the driven system has been
defined. Using the Baker-Campbell-Hausdorff (BCH) formula, one finds that $%
H_{\text{eff}}$ is formally given by
\begin{eqnarray}
H_{\text{eff}}&=& \frac{H_1}{2}+ \frac{H_2}{2} - \frac{iT}{8\hbar} [H_2, H_1]
\notag \\
&&\ - \frac{T^2}{96\hbar^2}\left[(H_2-H_1), [H_2,H_1]\right] + \cdots\ .
\label{bch}
\end{eqnarray}
Clearly then, even if $H_1$ or $H_2$ are short-range Hamiltonians,
the engineered $H_{\text{eff}}$ may still have long-{range} hopping
or pairing terms via the nested-commutator terms in Eq.~(\ref{bch}).
This constitutes a main difference from undriven systems. Thus, the
remaining job is to design such a protocol so that $H_{\text{eff}}$
also possesses time-reversal symmetry. Interestingly, in the first
proposal to realize Floquet MMs,~\cite{Jiang2011} {no more than two
pairs} of MMs can be generated precisely because time-reversal
symmetry is not restored by the periodic driving therein.

In the following we present our detailed results using a model of a 1D
spinless $p$-wave superconductor with the nearest- and next-nearest-neighbor
(NNN) interactions only. 
Under a periodic modulation of superconducting phases, we not only
demonstrate that many Floquet MMs (e.g. 13 pairs in one case) can be
generated, but also show that the number of the MMs may be widely
tuned by scanning the modulation period. These results also shed
more light on the inherent advantages of driven systems in exploring
new topological states of matter, which can be useful for other
timely topics related to long-range interactions (e.g., fractional
Chern insulators. \cite{Fractional2011,Fractional2012})

\section{Static model}
We start from the Kitaev model Hamiltonian for a 1D spinless $p$-wave
superconductor
\begin{eqnarray}
H=-\mu\sum_{l=1}^{N}c_{l}^{\dag}c_{l}-\sum_{a=1}^2%
\sum_{l=1}^{N-a}(t_{a}c_{l}^{\dag}c_{l+a}+\Delta_{a}c_{l}^{\dag}c_{l+a}^{%
\dag}+\text{h.c.}),  \label{Hamilton} \nonumber
\end{eqnarray}
where $\mu$ is the chemical potential, $t_{a}$ and $\Delta_{a}=\left\vert%
\Delta_{a}\right\vert e^{i\phi_a}$ with $a=1$ ($a=2$) describes the nearest-
(next-nearest-) neighbor hopping amplitude and pairing potential
respectively, and $\phi_{a}$ is the associated superconducting phases. All
energy-related parameters are scaled by $|\Delta_1|$ and $\hbar=1$ is set in
our calculations.  Majorana operators here refer to $(c_l+c_l^{\dag})$ or $%
i(c_l-c_l^{\dag})$. Such synthesized MMs may appear as edge modes under open
boundary condition, if the bulk band structure is topologically nontrivial.

The relative phase $\phi=\phi_1-\phi_2$ determines the topological class of $%
H$. \cite{ryuNJP2010} For $\phi=0$ and $\pi$, $H$ has time-reversal
and particle-hole symmetries. These cases then belong to the
so-called ``BDI" class characterized by a topological invariant $Z$.
For other values of $\phi $, $H$ has particle-hole symmetry only and
falls into the so-called ``D" class characterized by a topological
invariant $Z_2$. The D class can generate at most one pair of MMs.
As to the BDI class, despite its potential in forming many MMs,
\cite{Schnyder2008} at most two pairs of MMs can be generated here
due to the short-range nature of $H$.

\section{Driven model}
We now turn to periodically driven cases under a protocol given by Eq.~(\ref%
{bf}). The emergence of Floquet MMs is directly connected to topological
properties of the eigenstates of the Floquet operator $U(T)$. Let $|u\rangle$
be an eigenstate of $U(T)$ with an eigenvalue $e^{-i\epsilon T}$, namely $%
U(T)|u\rangle = e^{-i\epsilon T}|u\rangle$. Evidently, the eigenvalue index $%
\epsilon$ is defined only up to a period $2\pi/T$ and hence called
``quasi-energy". The periodicity in $\epsilon$ may lead to a novel
topological structure in driven systems, with the corresponding
topological classification revealed by the homotopy groups.
\cite{Kitagawa2010} However, if the driven system belongs to a
trivial class to this novel topological structure, then topological
properties of the driven system is fully
characterized by $H_{\text{eff}}$ defined in Eq.~(\ref{bf}). \cite%
{Schnyder2008} This will be the case for our driving protocol
proposed below.
\begin{figure}[tbp]
\begin{center}
\includegraphics[width=0.95\columnwidth]{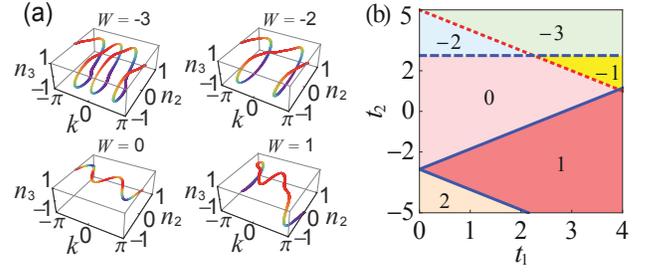}
\end{center}
\caption{(Color online) (a) Winding of the $\vec{n}(k)$ [see Eqs.~(\protect
\ref{n2eq}) and (\protect\ref{n3eq})] for $k\in [-\pi,\pi]$. $W=-3,-2,0,1$ correspond to $%
(t_1,t_2)$=$(1,5),~$(1,3),~$(1,0),~(1,-3)$, respectively. Indicated
on each panel is the winding number $W$. The solid and dotted line
indicates a gap closing (of $E_k$) at $k=0$ or $\pm \protect\pi$, while the
dashed line corresponds to a gap closing at $k=\protect\pi/2$. Other
parameters are $\protect\mu=-10$, $\left\vert\Delta_{2}\right\vert=2.5$ and $%
T=0.2$.}
\label{1st}
\end{figure}

As an explicit example, we propose to switch between two
Hamiltonians $H_1$ and $H_2$ by the following: in the first half
period, $H_1=H(\phi_1,\phi_2)$ with both superconducting phase
parameters $\phi_1$ and $\phi_2$ fixed; whereas in the
second half period, we swap $\phi_1$ and $\phi_2$ so that $%
H_2=H(\phi_2,\phi_1)$. Without loss of generality, we choose $%
\phi_1=\pi/2$ and $\phi_2=0$. Thus, within each half period, the
Hamiltonian is in class D that breaks time-reversal symmetry. In
addition to a possible generation of long-{range} interactions for
$H_{\text{eff}}$, this driving protocol is designed to recover
time-reversal symmetry. In particular, let ${\mathcal{K}}$ be a
conventional time-reversal operator and $G\equiv
e^{-i\frac{\phi_1{+}\phi_2}{2}\sum_{l}c_{l}^{\dag}c_{l}}$  be a
gauge transformation operator. Considering a generalized
time-reversal operator $\bar{\mathcal{K}}\equiv \mathcal{K}G$, we
find
\begin{eqnarray}
\bar{{\mathcal{K}}}U(T)\bar{{\mathcal{K}}}^{-1}= e^{\frac{iH_1 T}{2\hbar}}
e^{\frac{i H_2 T}{2\hbar}} = U^{\dag}(T).
\end{eqnarray}
This constitutes a direct proof that our driven system now possesses
time-reversal symmetry, and as a result its topological class is switched
from class D to class BDI. To further examine this restored time-reversal
symmetry, we work in the momentum representation and directly find an
analytical $H_{\text{eff}}$ from Eq.~({\ref{bf}). We define $c_{k}=
\sum_{l}c_{l}e^{- ikl}/\sqrt{N}$ and introduce the Nambu representation $%
C_{k}=[c_{k},c_{-k}^{\dag}]^{T}$. A standard procedure then leads to $H_{%
\text{eff}}=\sum_{k\in \text{BZ}}C_k^\dag H_{\text{eff}}(k)C_k$, with $H_{%
\text{eff}}(k)=E_k\vec{n}(k)\cdot\vec{\sigma}$, where $\vec{\sigma}$
represents the Pauli matrices. \cite{Kitagawa20101} The three components of $%
\vec{n}(k)$ are given by $n_1(k)=0$, and
\begin{eqnarray}
n_2 (k)&=&{\frac{g_{1,k}\sin(s_kT)}{s_k\sin (E_kT)}}-\frac{2g_{2,k}{\eta}%
_k\sin^2 (s_kT/2)}{s_k^2\sin (E_kT)},  \label{n2eq} \\
n_3 (k)&=&{\frac{{\eta}_k\sin(s_kT)}{s_k\sin (E_kT)}}+\frac{%
2g_{1,k}g_{2,k}\sin^2 (s_kT/2)}{s_k^2\sin (E_kT)},  \label{n3eq}
\end{eqnarray}
where $g_{a,k}=\left\vert\Delta_{a}\right\vert\sin (ak)$, $%
s_k=(\eta^2_k+\sum_a g_{a,k}^2)^{1/2}$, $\eta_{k}=-\mu-2\sum_{a}t_a\cos(ak)$%
,  and $\cos (E_kT)=\cos(s_kT) +2(g_{2,k}^2/s_k^2)\sin^2 (s_kT/2)$. For each
value of $k$, one obtains two values of $E_k$ and hence two values for the
quasi-energy $\epsilon$. Consistent with the $\bar{\mathcal{K}}$ symmetry,
we now have $H_{\text{eff}}^{*}(-k)=H_{\text{eff}}(k)$. Noting the inherent
particle-hole symmetry of $H_{\text{eff}}$, one may construct a chiral
symmetry for $H_{\text{eff}}$, a fact consistent with our above result that $%
\vec{n}(k)$ is in the $yz$ plane for all $k$. The above analysis
makes it clear that our driving protocol changes both the underlying
symmetry and the topological class of the system.

Without a gap closing between the two branches of $E_k$, the topological
invariant $Z$ in class BDI can be obtained by the integer winding number $%
W=\int_{-\pi}^{\pi}\frac{d\theta_{k}}{2\pi}\in Z$, where
$\theta_{k}=\arctan [n_3 (k)/n_2 (k)]$. A computational example
illustrating $W$ is shown in Fig.~\ref{1st}(a). The number of pairs
of MMs under open boundary condition is then given by $|W|$. As some
system parameters continuously change, gap closing and consequently
topological phase transitions occur. \cite{Read2000} Figure
\ref{1st}(b) depicts a phase diagram, obtained by explicitly
evaluating $W$. It is seen that $|W|$ ranges from 0 to $3$.
This indicates that three pairs of MMs can be formed in our driven
system. This is beyond the expectation for the undriven model, where
the NNN interaction can give at most two pairs of MMs. Therefore,
the finding of $|W|=3$ in some parameter regime is the first clear
sign that our driving protocol may synthesize some features absent
in the static model. The boundaries between different topological
phases of our driven system are also interesting on their own right.
The solid and dotted lines in Fig. \ref{1st}(b) depict the
topological phase transition points at which $W$ jumps by one. This
is found to go with the gap closing at $k=0$ or $\pm\pi$. The dashed
line gives the phase transition points at which $W$ jumps by two.
This happens at $k=\pi/2$.

\begin{figure}[tbp]
\begin{center}
\includegraphics[width=0.95\columnwidth]{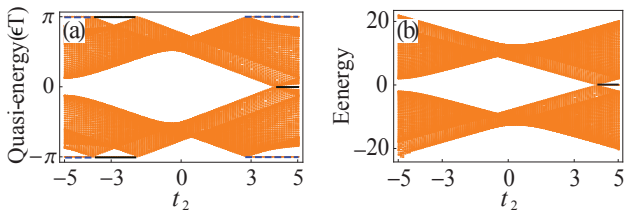}
\end{center}
\caption{(Color online) Quasi-energy spectrum for a
driven case (a) vs energy spectrum for a static case (b) obtained under
open boundary condition. Dashed blue line and solid black line stand for
two degenerate pairs and a single pair of MMs, respectively. $t_1=1$, {$N=200
$}, and other parameters are the same as in Fig. \ref{1st}(b). }
\label{bst}
\end{figure}

To confirm our theoretical results presented in Fig.~\ref{1st} we carry out
numerical calculations of the quasi-energy spectrum $\epsilon$ under open
boundary condition. Because $\epsilon=\pi/T$ is equivalent to $%
\epsilon=-\pi/T$, Floquet MMs have two flavors: one at $\epsilon=0$
and the other at $\epsilon=\pm\pi/T$. The second flavor is certainly
absent in an undriven system. \cite{Jiang2011} For fixed $t_1=1$ and
a varying $t_2$, Fig.~\ref{bst}(a) depicts the formation of both
flavors of Floquet MMs, with the second flavor emerging in a wider
parameter regime. The total number of pairs of MMs should equal
$|W|$ (if the winding number is well defined). For
example, Fig.~\ref{bst}(a) shows that two degenerate pairs of MMs at $%
\epsilon=\pm\pi/T$ and one pair of MMs at $\epsilon=0$ are formed when $t_1=1
$ and $t_2=4$. This agrees with the $W=-3$ region shown in Fig.~\ref{1st}%
(b). Likewise, all other details in Fig.~\ref{bst}(a) are fully consistent
with our analytical results shown in Fig.~\ref{1st}(b). We have also studied
the dynamics of the formed MMs in one full period of driving: they are
indeed well localized at two edges. Further, as a comparison with our static
model $H$, we plot in Fig. \ref{bst}(b) our system's energy spectrum in the
absence of driving. It is seen that at most one pair of MMs can be formed
only in a very narrow $t_2$ regime for the large $|\mu|$ case. The parallel
driven case is however different: one may still obtain three pairs of MMs.
Thus, even in the large $|\mu|$ case, our driving protocol can still
generate more MMs than the static case. This is both interesting and useful
because in general, the large $|\mu|$ is preferred for the protection of MMs
against strong disorder in actual experiments.
\begin{figure}[tbp]
\begin{center}
\includegraphics[width=0.95\columnwidth]{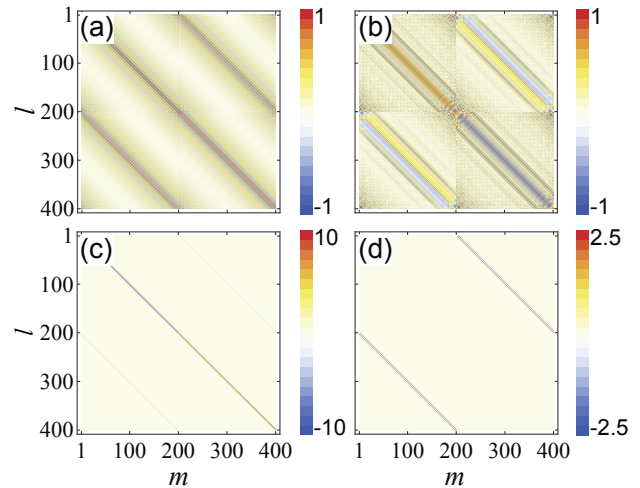}
\end{center}
\caption{(Color online) The expansion coefficients of $H_{\text{eff}}$ for $%
T = 0.2$ (a) and $2.0$ (b), and of the static $H$ for the real (c) and
imaginary (d) parts in the operator basis $(c_1,\cdots,c_N,c_1^\dag,%
\cdots,c_N^\dag)^T$. $l$ and $m$ are the base indices. Other parameters
are the same as in Fig.~\ref{bst}.}
\label{Hmatrix}
\end{figure}

In efforts to generate even more MMs, we now extend our direct numerical
studies to other parameter regimes. Remarkably, the BCH formula in Eq.~(\ref%
{bch}) indicates that as $T$ increases, the nested commutators on the right
hand side of Eq.~(\ref{bch}) will have heavier weights. An increasing $T$
can then induce longer-{range} interactions in $H_{\text{eff}}$. This trend
is investigated in Fig.~\ref{Hmatrix}, where the expansion coefficients of $%
H_{\text{eff}}$ [numerically obtained from Eq.~(\ref{bf})], with $H_{\text{%
eff}}$ expanded as a quadratic function of the operators $%
(c_1,\cdots,c_N,c_1^\dag,\cdots,c_N^\dag)^T$, are shown for two different
values of $T$. For comparison, the expansion coefficients for the static
case are also plotted in Fig. \ref{Hmatrix}(c,d). A few interesting
observations can be made from Fig.~\ref{Hmatrix}. First, the plotted
expansion coefficients of $H_{\text{eff}}$ are all real, which is different
from the shown static case with both real and imaginary coefficients. This
difference reflects the restored time-reversal symmetry for the driven case.
Second, in sharp contrast to the results shown in Fig. \ref{Hmatrix}(c,d),
coefficients for quite long-range hopping/pairing (e.g., across more than 10
sites) can be appreciably nonzero for $H_{\text{eff}}$ in both cases of $T =
0.2$ and $T=2.0$. The latter case, plotted in Fig. \ref{Hmatrix}(b) with
wider stripes, confirms the emergence of longer-range terms with
considerable weights as $T$ increases. Third, the diagonal terms in the
expansion shown in Fig.~\ref{Hmatrix}(a,b) (which can be understood as an
effective chemical potential) are much smaller than the diagonal elements,
i.e. $|\mu|$, in Fig.~\ref{Hmatrix}(c). This further explains why a driven
system may generate many MMs despite a large $|\mu|$ in the undriven model.

\begin{table}[tbp]
\caption{Number of MMs localized at each boundary for different $T$. Other
parameters are the same as in Fig. \ref{bst}(a). }
\label{perd}\setlength{\tabcolsep}{0.8pt}
\par
\begin{center}
\begin{tabular}{|c|c|c|c|c|c|c|c|c|c|c|c|c|c|c|c|c|c|}
\hline\hline
$t_2$ & $-8$ & $-7$ & $-6$ & $-5$ & $-4$ & $-3$ & $-2$ & $-1$ & $~0~$ & $~ 1~
$ & $~2~$ & $~ 3~$ & $~ 4~$ & $5$ & $6$ & $7$ & $8$ \\ \hline
$T=0.5$ & $2$ & $4$ & $4$ & $3$ & $3$ & $2$ & $0$ & $0$ & $0$ & $1$ & $1$ & $%
2$ & $2$ & $4$ & $4$ & $4$ & $3$ \\ \hline
$T=1.0$ & $6$ & $6$ & $7$ & $7$ & $6$ & $3$ & $3$ & $2$ & $1$ & $1$ & $2$ & $%
5$ & $5$ & $6$ & $7$ & $7$ & $6$ \\ \hline
$T=2.0$ & $13$ & $13$ & $12$ & $11$ & $9$ & $8$ & $8$ & $1$ & $1$ & $3$ & $4$
& $7$ & $11$ & $10$ & $13$ & $13$ & $12$ \\ \hline\hline
\end{tabular}%
\end{center}
\end{table}

Results in Fig.~\ref{Hmatrix} motivate us to explore the formation
of Floquet MMs with sufficiently large values of $T$. There is also
one twist as we increase $T$. That is, the quasi-energy
gap may be generically closed at $\epsilon=0$.
Consequently the winding number $W$ is no longer well-defined. To characterize the topological phases at
$\epsilon=\pm\pi/T$, where MMs can still be topologically protected, we resort to another topological
invariant \cite{Sato2011} \begin{equation}\nu=\frac{1}{2}\sum_{n_3(k)=0,E_{k}\neq0}\text{sgn}\{\partial_{k}[E_kn_3(k)]\}\text{sgn}[E_kn_2(k)],
\end{equation} which reduces to the winding number $W$ when the gap at $\epsilon=0$ is also open. We present in
Tab.~\ref{perd} the number of pairs of MMs we obtain, for an
increasing $T$ and for different choices of $t_2$. For $T=0.5$, the
best observation is the generation of $4$ pairs of MMs but no MMs
for $|t_2| \lesssim |t_1|$. For $T=1.0$, it is possible to achieve
$7$ pairs. For $T=2.0$, as many as 13 pairs of MMs can be formed.
Interestingly, in cases with large $T$ such as $T=1.0$, our driving
protocol can also form several pairs of MMs for $|t_2|<|t_1|$ or
even with $t_2=0$. Note that as more MMs are generated by increasing
$T$, the bulk quasi-energy gap at $\varepsilon=\pm\pi/T$ decreases
in general, leading to a larger ``penetration length" (into the
bulk) for the synthesized MMs. Considering the necessary protection
of MMs by a nonzero bulk gap, one may not wish to push our driving
protocol too far.

It is noted that the finite switching-time in the practise to our ideal step-driving scheme has no qualitative change to our results. However, it may influence quantitatively the range of the synthesized interaction as well as the numbers of the generated MMs. To simulate the smooth switching we have separated each of the two half-periods of our driving scheme into fifteen intermediate staircase-like changes and confirmed numerically that longer range interactions as well as more MMs can be generated. As a final remark, the number of the MMs characterized by the topological invariant depends on the topological properties of the Floquet states, which are determined by all the physical parameters in the driven model.

\section{Conclusions}
A periodic driving has the capacity to restore time-reversal symmetry and to
induce an effective long-{range} interaction. With these two mechanisms
working at once, the generation of many MMs is achieved using a standard $p$%
-wave superconductor model under certain periodic modulation.

In terms of possible experimental confirmation of our predictions,
our model may be realized with cold atoms or molecules in a designed
optical lattice, as clean systems with negligible perturbations. Explicitly, the nearest and NNN hopping ($t_a$) can
be realized by a simple zigzag chain lattice, \cite{Kraus2012} with
the hopping strength adjustable by the lattice geometry. The
chemical potential ($\mu$) is controllable through the optical trap
potential or a radio frequency detuning. The pairing terms
($\Delta_a$) may be induced by a Raman induced dissociation of
Cooper pairs forming an atomic BCS reservoir and the associated
superconducting phases ($\phi_a$) can be tuned by complex Rabi
frequencies. \cite{Jiang2011} Another experimental realization is to
use the recently proposed quantum-dot-superconductor arrays with a
zigzag geometry. \cite{Sau2012} Here $\mu$ can be gate controlled.
$\Delta_a$ can be proximity-induced and $\phi_a$ can be tuned via
applying fluxes on the superconducting islands. The MMs formed in
our system may be probed using techniques analogous to what is being
used for undriven systems, \cite{Kraus2012} but now with the hope of
some enhanced signals if a measurement exploits the simultaneous
generation of \textit{many} MMs. The generation of a tunable number
of many MMs is also expected to offer a new dimension for
experimental studies.

\section{Acknowledgements}
This work is supported by the Fundamental Research Funds for the Central
Universities, by the NSF of China (Grant Nos. 11175072, 11174115, and
10934008), and by National Research Foundation and Ministry of Education,
Singapore (Grant No. WBS: R-710-000-008-271). J.G. was funded by Academic
Research Fund Tier I, Ministry of Education, Singapore (grant No.
R-144-000-276-112).

\end{document}